\documentstyle[preprint,revtex,eqsecnum]{aps}
\begin{document}

\begin{title}
Resonances in $\Lambda d$ Scattering and the $\Sigma$-hypertriton
\end{title}

\author{I.~R.~Afnan}
\begin{instit}
  School of Physical Sciences, The Flinders University of South Australia,\\
  Bedford Park, S. A. 5042, Australia,\\
  and\\
  Institute for Nuclear Theory, University of Washington\\
  Seattle, Washington 98195
\end{instit}

\author{B.~F.~Gibson}
\begin{instit}
Theoretical Division,
Los Alamos National Laboratory, \\
Los Alamos, New Mexico 87545,\\
  and\\
  Institute for Nuclear Theory, University of Washington\\
  Seattle, Washington 98195
\end{instit}

\begin{abstract}
Using separable $NN$ and $\Lambda N$-$\Sigma N$ potentials in the Faddeev
equations, we have demonstrated that the predicted enhancement in the
$\Lambda d$ cross section near the $\Sigma d$ threshold is associated with
resonance poles in the scattering amplitude.  The positions of these poles,
on the second Riemann sheet of the complex energy plane, are determined by
examining the eigenvalues of the kernel of the Faddeev equations. This
suggests that for a certain class of $\Lambda N$-$\Sigma N$ potentials we can
form a $\Sigma$-hypertriton with a width of about 8 MeV.
\end{abstract}


\narrowtext

\section{INTRODUCTION}\label{sec:1}

In the realm of nonperturbative Quantum Chromo Dynamics (QCD) our description
of nuclear phenomena in terms of the physically observable baryons and mesons,
the collective modes of the QCD Lagrangian, has enjoyed considerable success.
A nonrelativistic two-body potential model picture of the $^3$H, $^3$He, and
$^4$He bound states as well as few-nucleon low-energy scattering and reactions
accounts amazingly well for much of the data.  The addition of the strangeness
degree of freedom to the nucleus opens the opportunity to ascertain whether
these models have predictive power or are merely vehicles of interpolation.
That is, can one use the models which have been developed in the
conventional, zero strangeness sector to extrapolate beyond that domain
to understand the nuclear physics involving $\Lambda$s and $\Sigma$s?

Although that question remains open, the strong coupling of the
$\Lambda N$-$\Sigma N$ system has been seen to lead to the enhancement of
certain phenomena which appear in nonstrange nuclei.  For example,
three-body-force effects in the binding energy of the hypertriton
($^3_\Lambda$H), when one eliminates the $\Sigma N$ channel from
the problem, are significant~\cite{AG90}; {\it i.e.},
$\Lambda N$-$\Sigma N$ coupling effects in hypernuclei appear
to play a much larger role than do $NN$-$\Delta N$ coupling
effects in nonstrange nuclei.  Furthermore, charge symmetry breaking, which
is strongly masked in the $^3$H$-^3$He isodoublet by the Coulomb force acting
between the two protons in $^3$He, is clearly obvious in the
$^4_\Lambda$H$-^4_\Lambda$He binding energy difference~\cite{Gib79}.  Thus,
extending our nuclear physics investigations to include $S \neq 0$ can
magnify certain physical effects.

While the existence of strangeness $-1$ $\Lambda$-hypernuclei is well
established from the observation of many bound states, such has not been the
case for $\Sigma$-hypernuclei.  Although structure in the recoilless production
of p-shell hypernuclei did indicate the possible existence of
$\Sigma$-hypernuclei~\cite{Ber80,Pie82,Yam85}, this structure corresponded
to unbound states.  Therefore, it was surprising to many when Hayano
{\it et al.}~\cite{Hay89} reported that the $\pi^-$ spectrum from the
$^4$He(stopped $K^-,\pi^{\pm}$) reactions exhibited narrow structure below
the threshold for $\Sigma$ emission.  It is the interpretation of such
spectra that we address in this investigation.

Charge conservation forbids the conversion of $\Sigma^- n$ into any $\Lambda N$
charge state.  If a $\Sigma^- n$ bound state were to exist, it would decay
to $\Lambda N$ only by the weak interaction.  Potential model analyses
of hyperon-nucleon ($YN$) scattering indicate a weak repulsion in the
spin-triplet state and nonbinding attraction in the spin-singlet state.
[The $\Sigma^- n$ system belongs to the same $SU(3)$ multiplet as the $nn$
system, which is almost bound in the $^1$S$_0$ state.]  The absence of
binding in the $\Sigma^- n$ system was confirmed by May
{\it et al.}~\cite{May82} through investigation of the
$^2$H($K^-,\pi^+)\Sigma^- n$ reaction.  However, this did not rule out
the possibility that the $\Sigma^- nn$ system might be bound. Such a bound
state would also be stable against $\Sigma N \rightarrow \Lambda N$
conversion.  However, in an analysis of the $\Sigma NN$ states, Dover and
Gal \cite{Dov82} noted that, if a bound state were to exist,
then the ($T=0$, $S=1/2$) configuration should lie lowest while the $T=2$
state would be the least likely to be bound, because of the spin-isospin
dependence of the $\Sigma N$ residual interaction.

An analogous analysis of the $A=4$ $\Sigma NNN$ system~\cite{Dov82} indicated
that the ($T=1/2$, $S=0$) configuration should lie lower in energy than the
($T=3/2$, $S=0$) configuration, although the latter state was expected to be
narrower.  Thus, the report by Hayano {\it et al.}~\cite{Hay89} that the
$\pi$ spectrum from stopped $K^-$ in the reaction $^4$He($K^-,\pi^-$)
exhibited narrow structure below the threshold for $\Sigma$ production was
quite exciting.  The ($K^-,\pi^-$) reaction can lead to both $T=1/2$
and $T=3/2$ $\Sigma NNN$ states, while the ($K^-,\pi^+$) reaction leads
only to the $T=3/2$ state.  Therefore, because no such structure was
observed in the spectrum from the complementary $^4$He($K^-,\pi^+$)
reaction, and because the ($K^-,\pi^-$) spin-flip reaction is small, the
structure was interpreted as a bound $^4_\Sigma$He state having the quantum
numbers $T=1/2$ and $J^{\pi}=0^+$.

Hayano has recently reported \cite{Hay91b} new results for in-flight
$^4$He($K^-,\pi^{\pm}$) experiments at BNL, which confirm the structure in
the ($K^-,\pi^-$) reaction and lack of structure in the ($K^-,\pi^+$)
reaction observed in the stopped $K^-$ absorption experiments.
The peak in the $\pi^-$ spectrum appears to be centered at
B$_{\Sigma^+} = 4 \pm 1$ MeV, consistent with the earlier result~\cite{Hay89}.
The width of the peak is about $10 \pm 2$ MeV, again consistent with a more
refined analysis of the KEK data~\cite{Hay91a}.  Furthermore, the data
are not inconsistent with the earlier bubble chamber data~\cite{Roo79} for
the exclusive $K^-$$^4$He$\rightarrow\pi^-\Lambda pd$ measurement,
recently reanalyzed by Dalitz {\it et al.}~\cite{Dal90}, which appear
to show a cusp-like enhancement near the $\Sigma^+$ production threshold.
The inferred $A=4$ $\Sigma-$hypernucleus would seem to
be more bound (by an MeV) than the $\Lambda$ is bound in $^4_\Lambda$He.
Although the $\Sigma$ is 10\% more massive than the $\Lambda$ which reduces
its kinetic energy, it would appear that the $\Sigma^+$ interaction
with $^3$H in the ($T=1/2$, $S=0$) channel must be more attractive than
the corresponding $\Lambda$ interaction with $^3$He or $^3$H.

Following the work of Dover and Gal on the ordering of the $A=4$
$\Sigma NNN$ states, Harada {\it et al.}~\cite{Har87,Har90} predicted
the existence of an $A=4$ $\Sigma NNN$ bound state using their
SAP-1 approximation to the Nijmegen YN potential model D~\cite{NIJ77}:
B$_{\Sigma^+} = 4.6$ MeV, $\Gamma = 7.9$ MeV.  They predicted no other bound
state for $A = 2-5$.  Nonetheless, we were motivated to examine the
$\Sigma$NN system in an effort to understand the properties of the
scattering amplitude with respect to observable structure in the physical
cross section.  For a sufficiently attractive $\Sigma N$ interaction, one
would hope to see evidence of a ($T=0$, $S=1/2$) $\Sigma NN$ bound state
or a low lying resonance in the $\Lambda d$
cross section near the threshold for $\Sigma$ production.  (Theoretical
models of the $YN$ interaction can exhibit a cusp phenomena in the $\Lambda N$
channel as one crosses the $\Sigma N$ threshold~\cite{Pea89}, but that
cusp dissolves into the continuum in the three-body system, where the
lowest threshold is not a two-body system but the $\Lambda d \rightarrow
\Sigma N N$ reaction channel.)  Although such model calculations are
not directly applicable to the $T=1$ inflight $^3$He($K^-,\pi^{\pm}$)
measurements that have been reported by Hungerford~\cite{Hun91} and
discussed by Hayano \cite{Hay91b}, they are relevant to the $T=0$
$^3$H($K^-,\pi^-$) reaction as well as to analysis of $\Lambda d$
scattering.

In this paper we explore the structure of the $\Lambda d$ cross
section in terms of a Hamiltonian model.  For Hermitian Hamiltonians
the spectrum consists of the eigenvalues for the bound and scattering
states.  From the scattering state eigenfunctions we can extract the
scattering amplitude and, therefore, the cross section.  The presence
of rapid fluctuations (structure) in the cross section is normally
attributed to resonances, which can be viewed as poles in the scattering
amplitude on the {\it second} Riemann sheet of the complex energy plane.
It is possible to establish a direct relation between the Hamiltonian
for the system and the resonance energies and widths by realizing that
the Hermitian Hamiltonian (and corresponding eigenvalue problem) is
defined on the {\it first} Riemann sheet of the complex energy plane,
while the poles of the scattering amplitude are on the {\it second}
Riemann sheet. Thus, to directly obtain the desired resonance energies
(poles), one must analytically continue the eigenvalue problem onto that
part of the second sheet where the resonance poles reside.  This leads
to an eigenvalue problem for a non-Hermitian Hamiltonian which,
therefore, admits complex eigenvalues.  These complex eigenvalues specify
the energy and width of the resonances.  The corresponding wave functions
are normalizable, as we shall see below, provided one realizes that the
solutions of a non-Hermitian eigenvalue problem and the definition of the
normalization must be appropriately modified.

In terms of the specific problem at hand, if the $YN$ interaction
produces a pole in the $\Lambda d$  amplitude below the $\Sigma NN$
threshold (and on the top sheet of the  $\Sigma NN$ branch cut but the
bottom sheet of the $\Lambda NN$ branch cut, the  $[bt]$ Riemann
sheet~\footnote{ We adopt the convention of Ref.~\cite{Pea89}  for the
labeling of the Riemann sheets corresponding to the $\Lambda NN$ and
$\Sigma NN$ threshold.  However, in this problem we have additional sheet
structures from the $\Lambda d$ threshold and any additional branch
points arising from the resonance poles of the $YN$ $t$ matrix.}), then
one would  anticipate narrow structure in $\Lambda d$ scattering below
the $\Sigma NN$  threshold.  In contrast, if the $YN$ interaction
produces a pole above  the threshold in the $\Sigma NN$ system (and on
the top sheet of the  $\Sigma NN$ branch cut but on the second sheet of
the $\Lambda NN$  branch cut, again the [bt] sheet), then the effect of
this pole will still be  to produce structure in the $\Lambda d$ cross
section below the  $\Sigma NN$ threshold.  This occurs because, for
energies above the  $\Sigma NN$ threshold, the pole is screened from the
physical region by the  branch cut due to the presence of the threshold.
To see structure above  the $\Sigma NN$ threshold, there should be a pole
on the second sheet of both the $\Lambda NN$ and $\Sigma NN$ branch cut,
{\it i.e.} $[bb]$, above the $\Sigma$ production threshold.  That is,  any
structure seen below the $\Sigma$ production threshold will be due to
the poles on the $[bt]$ Riemann sheet.  Such a pole might correspond to
(i) a  bound state of the $\Sigma NN$ system in the absence of coupling of
the $\Sigma N$ channel to the $\Lambda N$ channel (a pole shifted into the
complex plane resulting in the structure seen in $\Lambda d$ scattering
when the $\Lambda N$-$\Sigma N$ coupling is turned on), or (ii) an unbound
state of the $\Sigma NN$ system in the absence of coupling to the $\Lambda N$
channel (a pole which is moved onto the [bt] sheet when the coupling is
turned on).  In either case, enhancement in the $\Lambda d$ cross section
below the $\Sigma$ production threshold corresponds to an  {\it eigenstate}
of the $YNN$ system.

To explore this hypothesis, we present a detailed discussion of the
equations describing the $YNN$ system in the presence of a $YN$
($\Lambda N$-$\Sigma N$ coupled-channel) potential in the following section.
Formal solution of the three-body equations is outlined in the Appendix.
Numerical results for specific $YN$ potential models are presented in Section
III.  A discussion of the results and summary of our conclusions can be found
in Section IV.

\section{THEORY}\label{sec:2}

To establish the connection between the enhancement in the cross section for
$\Lambda d$ scattering and the formation of a $\Sigma$-hypertriton, we must
demonstrate that the structure found in the cross section for $\Lambda d$
scattering is due to poles in the scattering amplitude on the second energy
sheet, and that these poles correspond to eigenstates of the Hamiltonian for
the $YNN$ system in which $Y=\Lambda,\ {\rm or}\ \Sigma$. This connection
between the cross section and the eigenstates of the Hamiltonian is achieved
by: (i)~showing formally that the energy at which the scattering amplitude
has a pole on the second energy sheet can be considered an eigenstate of
the Hamiltonian, (ii)~demonstrating that for the specific models of the
$\Lambda N$-$\Sigma N$ interaction considered, there is a correlation between
the enhancement in the cross section and the position of the poles of the
scattering amplitude, or $T$-matrix.

To establish the fact that the position of a pole in the scattering amplitude
corresponds to an eigenstate of the Hamiltonian, we consider the $YNN$ system
in terms of a three-body Hamiltonian given by
\begin{equation}
H = H_0 + V  \ ,                                       \label{eq:2.1}
\end{equation}
where $H_0$ is the kinetic energy of the three-particle system and $V$ is
the sum of pairwise interactions. In spectator particle notation $V$ is given
by
\begin{equation}
V = \sum_{\alpha=1}^3\ V_\alpha\ ,                     \label{eq:2.2}
\end{equation}
with $V_3$ being the $NN$ interaction, while $V_1$ and $V_2$ are the $YN$
interactions. The Schr\"odinger equation for this three-body system can
then be written as
\begin{equation}
\left(E - H_0\right)\,|\Psi\rangle = V\,|\Psi\rangle\ , \label{eq:2.3}
\end{equation}
or
\begin{eqnarray}
|\Psi\rangle &=& G_0(E)\,V\,|\Psi\rangle \nonumber \\
&=& \sum_{\alpha=1}^3\ G_0(E)\,V_\alpha\,|\Psi\rangle \nonumber \\
&\equiv&\sum_{\alpha=1}^3\ |\psi_\alpha\,\rangle \ ,      \label{eq:2.4}
\end{eqnarray}
where the free Green's function $G_0(E) = (E-H_0)^{-1}$. The last line in
Eq.~(\ref{eq:2.4}) corresponds to the Faddeev decomposition of the wave
function. The Faddeev components of the wave function $|\psi_\alpha\,\rangle$,
then satisfy the equation
\begin{equation}
|\psi_\alpha\,\rangle = \sum_\beta\ G_0(E)\,V_\alpha\,|\psi_\beta\,\rangle\ ,
                                                      \label{eq:2.5}
\end{equation}
or
\begin{equation}
\left[1 - G_0(E)\,V_\alpha\right]\,|\psi_\alpha\,\rangle =
\sum_\beta\,G_0(E)\,V_\alpha\,\bar{\delta}_{\alpha\beta}\,|\psi_\beta\,\rangle
\ ,                                                   \label{eq:2.6}
\end{equation}
where $\bar{\delta}_{\alpha\beta} = (1 -\delta_{\alpha\beta})$. If we now
multiply
Eq.~(\ref{eq:2.6}) by $[1 - G_0(E)\,V_\alpha\,]^{-1}$, and take into
consideration the fact that the $T$-matrix for the two-body sub-system,
$T_\alpha(E)$, is given by $T_\alpha(E) =
[1 - V_\alpha\,G_0(E)]^{-1}\,V_\alpha$, we can write the equation
for the Faddeev component of the wave function as
\begin{equation}
|\psi_\alpha\,\rangle = \sum_\beta\
G_0(E)\,T_\alpha(E)\,\bar{\delta}_{\alpha\beta}\, |\psi_\beta\,\rangle\ ,
                                                      \label{eq:2.7}
\end{equation}
or
\begin{equation}
|\phi_\alpha\,\rangle = \sum_\beta\ G_0(E)\,\bar{\delta}_{\alpha\beta}\,
T_\beta(E)\,|\phi_\beta\,\rangle\ ,                     \label{eq:2.8}
\end{equation}
where
\begin{equation}
|\phi_\alpha\,\rangle = \sum_\beta\ \bar{\delta}_{\alpha\beta}\,
|\psi_\beta\,\rangle\ .                                   \label{eq:2.9}
\end{equation}
It is the solution of Eq.~(\ref{eq:2.7}) or Eq.~(\ref{eq:2.8}), that gives the
bound state of the hypertriton. To that extent, the solution of either
Eqs.~({\ref{eq:2.7}) or (\ref{eq:2.8}) is identical to the solution of the
Schr\"odinger equation.  In fact, the energy at which these equations have
a solution corresponds to a bound state, and the solution is an eigenstate
of the Hamiltonian for the three-body system. In momentum representation this
homogeneous integral equation, Eq.~(\ref{eq:2.8}), has the same kernel as the
Alt-Grassberger-Sandhas (AGS) equations~\cite{Alt67} for three-particle
scattering, which we can write as,
\begin{equation}
 X_{\alpha\beta}(E) = \bar{\delta}_{\alpha\beta}\,G_0 +  \sum_\gamma \,
G_0(E)\,\bar{\delta}_{\alpha\gamma}\,T_\gamma(E)\,X_{\gamma\beta}\ .
                                                       \label{eq:2.10}
\end{equation}
This suggests that if we convert the homogeneous integral equation
Eq.~(\ref{eq:2.8}) to an eigenvalue problem of the form
\begin{equation}
\lambda_n(E)\,|\phi_{n,\alpha}\,\rangle =
\sum_\beta\ G_0(E)\,\bar{\delta}_{\alpha\beta}\,
T_\beta(E)\,|\phi_{n,\beta}\,\rangle   \ ,             \label{eq:2.11}
\end{equation}
where the $\lambda_n(E)$ are the eigenvalues and $|\phi_{n,\alpha}\,\rangle$
are the eigenstates, then the solution of the inhomogeneous integral equation,
Eq.~(\ref{eq:2.10}), for the amplitude $X_{\alpha\beta}$ can be written in
terms of the eigenvalues and eigenstates of the homogeneous equation,
Eq.~(\ref{eq:2.11}), as (see Appendix)~\cite{Afn91}:
\begin{equation}
X_{\alpha\beta}(E) = \sum_n\ |\,\phi_{n,\alpha}(E)\,\rangle\,
\frac{\left[\tilde{\lambda}_n(E^*)\right]^*}{1 - \lambda_n(E)}\,
\langle\,\tilde{\phi}_{n,\beta}(E^\ast)\,| \ .             \label{eq:2.12}
\end{equation}
Here, $|\tilde{\phi}_{n,\beta}\,\rangle$ and $\tilde{\lambda}_n$ are the
eigenstates and eigenvalues of the adjoint kernel. It is clear from
Eq.~(\ref{eq:2.12}) that for energies at which $\lambda_n(E)=1$, the
scattering amplitude $X_{\alpha\beta}(E)$ has a pole. Thus, the positions
of the poles of $X_{\alpha\beta}(E)$ on the second Riemann sheet
of the energy plane can be determined by examining the eigenvalues
of Eq.~(\ref{eq:2.11}) for complex energies.

Since resonance poles reside on the second Riemann sheet of the complex energy
plane, we deform our contour of integration in momentum space in order
to analytically continue our eigenvalue equation, Eq.~(\ref{eq:2.11}), onto
the second sheet.  However, the deformation of the contour of integration
requires a knowledge of the position of the singularities of the kernel in the
energy variable. In fact, as we will demonstrate, these singularities
constrain the energy domain onto which we can analytically continue our
equations. The singularities of the kernel of the AGS equation are determined
by the dynamics of the two-body interaction we include in our analysis.
Since we will restrict our calculations to separable two-body potentials
that include $\Lambda N$-$\Sigma N$ coupling, we can rewrite
Eqs.~(\ref{eq:2.10}) and (\ref{eq:2.11}) for this class of interactions.
These separable potentials can be written in matrix form, after partial wave
expansion, as~\cite{AG90}
\begin{equation}
V_\alpha = |\,g_{\kappa_\alpha}\,\rangle\,C_{\kappa_\alpha}\,
\langle\,g_{\kappa_\alpha}\,|  \ ,                         \label{eq:2.13}
\end{equation}
where $C_{\kappa_\alpha}$ is the strength of the interaction in the
$\kappa_\alpha$ partial wave, while $|g_{\kappa_\alpha}\rangle$ is the
corresponding form factor. The $t$ matrix, in two-body Hilbert space, for this
potential is then given by
\begin{equation}
t_{\kappa_\alpha}(\varepsilon_\alpha) = |\,g_{\kappa_\alpha}\,\rangle\,
\tau_{\kappa_\alpha}(\varepsilon_\alpha)\,\langle\,g_{\kappa_\alpha}\,|\ ,
                                                  \label{eq:2.14}
\end{equation}
where the ``quasi-particle'' propagator, $\tau_{\kappa_\alpha}$, takes the form
\begin{equation}
\tau_{\kappa_\alpha}(\varepsilon_\alpha) = \left[C^{-1}_{\kappa_\alpha} -
\langle\,g_{\kappa_\alpha}\,|\,g_0(\varepsilon_\alpha)\,|\,g_{\kappa_\alpha}\,
\rangle \right]^{-1}\ .
\label{eq:2.15}
\end{equation}
Here, $g_0(\varepsilon_\alpha)$ is the two-body free Green's function for the
pair $(\beta\gamma)$.

\widetext
With the above results for the two-body $t$ matrix, we can proceed to write the
AGS equations, and the corresponding homogeneous equation for a given total
angular momentum $J$ and isospin $T$ as~\cite{AG90}
\begin{eqnarray}
X^{JT}_{k_\alpha;k_\beta}(q,q';E^+) &=&
Z^{JT}_{k_\alpha;k_\beta}(q,q';E^+)
\nonumber \\
&&\qquad
+ \sum_{k_\gamma}\,\int\limits_0^\infty\,dq''\,
K^{JT}_{k_\alpha;k_\gamma}(q,q'';E^+)\,
X^{JT}_{k_\gamma;k_\beta}(q'',q';E^+)\ ,               \label{eq:2.16}
\end{eqnarray}
\narrowtext
and
\begin{eqnarray}
\lambda_n(E)\,\phi_{n,k_\alpha}(q;E) &=&
 \sum_{k_\beta}\,\int\limits^\infty_0\,
dq'\,K^{JT}_{k_\alpha;k_\beta}(q,q';E) \nonumber \\
&&\qquad \times \phi_{n,k_\beta}(q';E)
\ ,                                                    \label{eq:2.17}
\end{eqnarray}
where the kernel of the integral equations is given by
\begin{eqnarray}
K^{JT}_{k_\alpha;k_\beta}(q,q';E) &=& Z^{JT}_{k_\alpha;k_\beta}(q,q';E)\,
\nonumber \\
&&\qquad \times \tau_{\kappa_\beta}\left[E-\varepsilon_\beta(q')
\right]\,q'^2\ .                                        \label{eq:2.18}
\end{eqnarray}
Here, $k_\alpha$ refers to the set of quantum numbers that define the partial
wave three-body channel with particle $\alpha$ the spectator, while
$\varepsilon_\alpha$ is the energy of the spectator particle. The partial
wave Born amplitude, $Z^{JT}_{k_\alpha;k_\beta}$, is given by
\begin{eqnarray}
Z^{JT}_{k_\alpha;k_\beta}(q,q';E) &=& \bar{\delta}_{\alpha\beta}\,
\langle\,TJk_\alpha q;g_{\kappa_\alpha}| \, G_0(E) \nonumber \\
&& \qquad \times |g_{\kappa_\beta};q'
k_\beta JT\,\rangle\ .
                                                              \label{eq:2.19}
\end{eqnarray}
An explicit expression for this Born amplitude has been given
previously~\cite{AG90}.

To analytically continue Eq.~(\ref{eq:2.17}) onto the second Riemann sheet of
the complex energy plane, we rotate the contour of integration; {\it i.e.}, we
make the transformation
\begin{equation}
q\rightarrow q\,e^{-i\theta}\ ,\qquad q'\rightarrow q'\,e^{-i\theta}\quad\mbox
{with}
\quad \theta>0\ .                                       \label{eq:2.20}
\end{equation}
This should, in principle, extend the energy domain over which
Eq.~(\ref{eq:2.17}) is defined to that part of the second energy plane for
which
$|\arg{E}|< 2\theta$. However, the singularities of the kernel put a constraint
on the range of values $\theta$ can assume. Since both $q$ and $q'$ in
Eq.~(\ref{eq:2.17})  are rotated by the same angle, the singularities of the
Born amplitude are such that the only constraint they place on $\theta$ is that
$\theta<\frac{\pi}{2}$~\cite{St77,Afn91}. This, for all practical purposes,
imposes no serious constraint on the energy domain to which we can extend our
equation in order to search for resonance poles. This leaves us with
the singularities of the ``quasi-particle'' propagator $\tau_\kappa$, which are
of  two kinds: (i)~simple poles due to two-body bound or resonance states,
and (ii)~square root branch points which give rise to the unitarity cuts in the
two-body subsystem.  The class (i) poles lead to branch points
in the three-body amplitude, which correspond to the thresholds
for the production of a bound or resonant pair.  The class (ii)
branch points give rise to thresholds in the three-body amplitude.  Both
types of singularities can be exhibited by writing the quasi-particle
propagator as
\begin{equation}
\tau_{\kappa_\alpha}\left[E-\varepsilon_{\kappa_\alpha}(q')\right] =
\frac{S_{\kappa_\alpha}\left[E-\varepsilon_{\kappa_\alpha}(q')\right]}
{E-\varepsilon_{\kappa_\alpha}(q')-\epsilon_{r_\alpha}}\ ,
                                                        \label{eq:2.21}
\end{equation}
where
\begin{equation}
\epsilon_{r_\alpha} = \left\{ \begin{array}{ll} M_{b_\alpha} &\quad
\mbox{for two-body bound states}\\
 M_{r_\alpha} -\frac{i}{2}\Gamma_{r_\alpha} &\quad \mbox{for two-body
resonances} \end{array}\right.\ .
                                                         \label{eq:2.22}
\end{equation}
Here, $M_{b_\alpha}$ is the mass of the two-body bound state\footnote{We have
defined our energy $E$ to include the mass of the two nucleons and the
$\Lambda$.}, {\it i.e}. $M_{b_\alpha} = m_\beta + m_\gamma - B$, with $B$
the two-body binding energy, while $M_{r_\alpha}$ and
$\Gamma_{r_\alpha}$ are the mass and full
width of the resonance in the two-body  subsystem. In Eq.~(\ref{eq:2.21}) the
function $S_{\kappa_\alpha}(\varepsilon)$ has square root branch points at
$\varepsilon_{\kappa_\alpha}=m_\beta + m_\gamma$, while the energy
denominator  has the poles of the quasi-particle propagator. For the
$YNN$ system,  the deuteron quasi-particle propagator
$\tau_d[E - \varepsilon_d(q)]$ has a pole at  the deuteron mass, while
for the $\Lambda N$-$\Sigma N$ interactions, $\tau_{\kappa_\alpha}$
has the $\Lambda N$ and $\Sigma N$ threshold. In addition,
some $YN$ potentials have a resonance pole in the $^3$S$_1$ channel near the
$\Sigma N$ threshold. In Fig.~\ref{Fig.1}, we illustrate the position of these
branch cuts in the three-body energy plane when the contour of rotation  is
$\theta$.

To determine how far we can analytically continue Eq.~(\ref{eq:2.17}) into the
complex plane, we must examine how the singularities of $\tau_\kappa$
effect the rotation of the contour of integration. Taking
\[
\varepsilon_{\kappa_\alpha}(q') = m_\alpha + \frac{q'^2}{2\mu_\alpha} \ ,
\]
where $m_\alpha$ is the mass of the spectator particle
$\alpha$ and $\mu_\alpha$ is the reduced  mass of the spectator
$\alpha$ with the pair $\beta\gamma$, we
see that the branch point from  $S_{\kappa_\alpha}$ is at
\begin{equation}
q' = \pm\sqrt{2\mu_\alpha\left(E-\sum_i m_i\right)}\ .
                                                        \label{eq:2.23}
\end{equation}
For a three-body resonance with energy $E$ ({\it i.e.},
$E=E_r -iE_i,\ E_i>0$), these branch points are in the fourth quadrant
of the $q'$-plane and at an angle of $\varphi_u$, where
\begin{equation}
\tan 2\varphi_u = \frac{E_i}{E_r-\sum_i m_i}\ .       \label{eq:2.24}
\end{equation}
Thus, in as far as these branch points are concerned, we need to take
$\theta>\varphi_u$ to avoid the singularities. On the other hand the
poles of $\tau_{\kappa_\alpha}$ are at
\begin{equation}
q' = \pm\sqrt{2\mu_\alpha\left(E - m_\alpha - \varepsilon_{r_\alpha}\right)}\ .
                                                    \label{eq:2.25}
\end{equation}
For the case of the deuteron bound state ({\it i.e.} $\varepsilon_{r_\alpha} =
m_\beta + m_\gamma - B$) the  quasi-particle propagator has poles in the fourth
quadrant of the  $q'$-plane at an angle $\varphi_d$, where
\begin{equation}
\tan2\varphi_d = \frac{E_i}{E_r+B-\sum_i m_i}\ .       \label{eq:2.26}
\end{equation}
Since $\varphi_d<\varphi_u$, we need not worry about this pole putting any
constraint on the contour rotation. That leaves us with the resonance poles
in the quasi-particle propagator for the $\Lambda N$-$\Sigma N$ interaction.
In this case the angle of the resonance pole in the $q'$-plane is
$\varphi_r$, where
\begin{equation}
\tan2\varphi_{r_\alpha} = -\frac{E_i - \frac{1}{2}\Gamma_{r_\alpha}}
{E_r -M_{r_\alpha} -m_\alpha}\ .                      \label{eq:2.27}
\end{equation}
For $E_r < (M_{r_\alpha}+m_\alpha)$ and $E_i < \frac{1}{2}\Gamma_{r_\alpha}$,
the
angle  $2\varphi_r$ is in the second quadrant, and therefore,
$\frac{\pi}{4}<\varphi_r<\frac{\pi}{2}$. As we proceed along the real axis
to the point $E=(M_{r_\alpha}+m_\alpha)$, $\varphi_r$ attains a value of
$\frac{\pi}{4}$, while proceeding parallel to the imaginary axis to the point
$E_i=\frac{i}{2}\Gamma_{r_\alpha}$, $\varphi_r$ attains a value of
$\frac{\pi}{2}$. If we carry this analysis through, we find that as
we analytically continue our equation in the energy variable from the
real axis through region $I$ to region $III$ and then to region $IV$
(see Fig.~\ref{Fig.2}), one of the resonance poles in the $q'$-plane
moves into the region $-\frac{\pi}{4}<\varphi_r<0$ approaching from
$\varphi_r=-\frac{\pi}{4}$. At this stage the two-body unitarity branch point
is moving towards $\varphi_u=\frac{\pi}{4}$. These two singularities
could force the contour to deviate from the path along the ray, and this in
turn will introduce logarithmic branch points from the Born
amplitude $Z^{JT}_{k_\alpha;k_\beta}$.
Thus, the energy domain on the second Riemann sheet, to which we can
analytically continue Eq.~(\ref{eq:2.17}) without introducing elaborate
contours of integration, is shown as the shaded area in
Fig.~\ref{Fig.3}~\cite{PA84}.  In addition to the above energy domain, we can
analytically  continue Eq.~(\ref{eq:2.17}) onto the third Riemann sheet through
the branch  cut generated by the resonance pole in $\tau_{\kappa_\alpha}$;
{\it i.e.}, we start on the real axis in region $II$, then proceed through
the branch cut to region $IV$ onto the third  Riemann sheet, and
then to region $III$ on the third Riemann sheet (see Fig.~\ref{Fig.2}).
In this case as we proceed from region $II$ to region $IV$,
the resonance pole in the $q'$-plane crosses the
real axis into the fourth quadrant, and we can analytically continue the
equation into region $IV$ and then $III$ of the third Riemann sheet. However,
if we attempt to go to region $I$ of the third energy sheet, we find that
the contour of integration is forced onto the negative imaginary $q'$-axis by
the two-body resonance pole, and here we encounter the singularities of the
Born amplitude. Thus, the only part of the third Riemann sheet of the
complex energy plane that we can access is the shaded region in
Fig.~\ref{Fig.4}. In the next section we will use the above results
to explore the region near the $\Sigma NN$ threshold for possible
resonances that might explain the structure we see in the cross section
for $\Lambda d$ scattering.

\section{NUMERICAL RESULTS}

To examine the possible existence of $\Sigma$-hypernuclear states below the
$\Sigma$-production threshold in the $A=3$ system, we must first consider
two-body interactions that could generate such resonances. In particular, we
need to know what features of the two-body interaction would produce a
resonance in the $YNN$ system. This is particularly important as the
hyperon-nucleon ($YN$) interactions we use are of separable potential
form, and with the limited data available such separable potentials are not
uniquely determined. Ideally, we would like to carry out the computational work
for the more realistic $YN$ interaction such as the One Boson Exchange (OBE)
potentials in which the extensive $NN$ and limited $YN$ data are considered
within the unified framework of $SU(3)$. However, this is not justifiable at
this stage, considering the lack of information about the correlation between
the results of the two-body $YN$ and three-body $YNN$ systems. Therefore, as a
first calculation we utilize several separable potentials previously employed
in light hypernuclei investigations.  We then vary the strength of the
coupling in the $^3$S$_1$ partial wave between the $\Lambda N$ and $\Sigma N$
channel to explore the variation in the two-body and three-body results.

\subsection{The Two-Body Input}

For the present calculations we restrict our $YN$ two-body interactions to
$S$-wave. For the $NN$ interaction we use the same
potentials previously used in our study of the role of $\Lambda N-\Sigma N$
coupling in the hypertriton~\cite{AG90}. In particular, we use a Yamaguchi
potential for the $^1$S$_0$, and the Phillips~\cite{Ph68} potential with
$P_D=4\%$ for the $^3$S$_1 - ^3$D$_1$ partial wave. The parameters of these
potentials, in the present notation, are given in Ref.~\cite{AG90}.

For the $YN$ interaction in the $^1$S$_0$ partial wave we use the potential
of Stepien-Rudzka and Wycech (SRW)~\cite{SRW81}. Here again the parameters of
this potential, in the present notation, were given previously in
Ref.~\cite{AG90}. Since the coupling between the $\Lambda N$ and $\Sigma N$
channel has a one pion exchange contribution, we expect the $^3$S$_1$
channel to be stronger in its long range behaviour than the corresponding
$^1$S$_0$. We therefore have chosen to vary the interaction in this partial
wave only. The potentials we have used are the coupled-channel SRW
potential, and the potentials constructed by Toker, Gal, and
Eisenberg~\cite{TGE81} (TGE). The latter potentials where constructed to
investigate the question of the possible existence of resonances in
$K^- d\rightarrow\pi N\Lambda$ near the $\Sigma$ threshold. In particular,
potentials B and C, to which we will refer as TGE-B and TGE-C respectively,
support a $\Sigma N$ bound state in the absence of coupling between
the $\Lambda N$ and $\Sigma N$ channels (see Table~\ref{table1}),
while potential A, referred to here as TGE-A, has a virtual state in
the absence of coupling. In Table~\ref{table1} we
present the parameters of these potentials, while in Table~\ref{table2} we give
the positions of the poles in the complex energy plane with and without the
coupling between the $\Lambda N$ and $\Sigma N$ channels. Included in the
tables
are also the parameters of the $^1$S$_0$ potential of SRW, and the position of
the poles for this potential. In Table~\ref{table2} we have used the notation
of
Pearce and Gibson~\cite{Pea89} for specifying the sheet on which the pole
resides. Thus,  $[tt]$ corresponds to the top sheet of both the $\Lambda N$ and
$\Sigma N$ branch cuts, while $[bt]$ corresponds to the bottom sheet of the
$\Lambda N$ branch cut and the top sheet of the $\Sigma N$ branch cut. In
Fig.~\ref{Fig.5} we illustrate the sheet labeling system for the $YN$ problem,
with two square root branch cuts corresponding to the $\Lambda N$ and $\Sigma
N$
thresholds. From Table~\ref{table2}, we observe that potentials TGE-B and
TGE-C have poles on the $[bt]$ sheet and in the absence of coupling between the
$\Lambda N$ and $\Sigma N$ channels these poles become $\Sigma N$ bound states.
In fact, as the coupling between the two channels changes these poles move
continuously, tracing a path on the $[bt]$ sheet. On the other hand, for the
potentials TGE-A and SRW the pole near the $\Sigma N$ threshold resides on the
$[tb]$ sheet. In this case, turning off the coupling brings the pole to the
real energy axis and on the second sheet of the $\Sigma N$ branch cut. This
corresponds to a virtual state of the $\Sigma N$ system. In particular, we
should note that for the SRW potential we have a zero energy bound state
in the absence of coupling.

Because we are considering two classes of potentials, those with a bound
$\Sigma N$ and those with a virtual, or unbound, $\Sigma N$ in the absence
of coupling between the two channels, one might like to compare at the same
time the effective ranges parameters for these potentials, and possibly
compare them to the more ``realistic'' OBE potentials. For that we would like
to
calculate the effective range parameters, and particularly the effective range
parameters in the $\Sigma N$ channel. These effective range parameters, which
will be complex for the $\Sigma N$ system, are defined in terms of the two-body
diagonal partial-wave $T$-matrix in channel $\alpha$ as
\begin{equation}
-\frac{1}{a_\alpha} + \frac{1}{2}k^2r_\alpha =
- \frac{1 - i\pi\mu_\alpha k_\alpha T_{\alpha\alpha}}{\pi\mu_\alpha
T_{\alpha\alpha}} ,                                      \label{eq:3.1}
\end{equation}
where the $T$-matrix is given in terms of the phase shifts by the
relation
\begin{equation}
T_{\alpha\alpha} = -\frac{1}{\pi\mu_\alpha k_\alpha}\,
e^{i\delta_\alpha}\sin{\delta_\alpha} \ .
                                                      \label{eq:3.2}
\end{equation}
Here, $k_\alpha$ is the on-shell momentum in a given channel, while
$\mu_\alpha$
and $\delta_\alpha$ are the reduced mass and phase shift in channel $\alpha$
respectively. In Table~\ref{table3} we present the effective range parameters
for the four $^3$S$_1$ potentials under consideration, and the $^1$S$_0$ SRW
potential. From this table we observe that potentials TGE-B and TGE-C have a
$\Sigma N$ scattering length with a positive real part, while for potential
TGE-A, which has a virtual state, the real part of the $\Sigma N$ scattering
length is negative. For potential SRW this simple one-channel interpretation of
the sign of the scattering length does not work. This suggests that we need to
examine the position of the poles of the scattering amplitude, which in general
are in the complex energy plane, before we can make any statement about whether
the $\Sigma N$ interaction supports a bound state.

Since we will examine the cross section for $\Lambda d$ scattering as a
means of determining the presence or absence of resonances, we should study
at the same time the cross section for $\Lambda N$ scattering in the $^3$S$_1$
channel, to investigate whether there are correlations between the results
for the two- and three-body systems. In particular, we would like to compare
the
case when the two-body $\Sigma N$ system supports a bound state in the absence
of $\Lambda N-\Sigma N$ coupling, verses the case when there is a virtual state
for the uncoupled $\Sigma N$ system\footnote{Here, we should remind the reader
that a bound state corresponds to a pole on the first sheet, while a virtual
state correspond to a pole on the second sheet of the energy plane.}. Finally,
we would like to investigate whether the shape of the cross section provides
any indication as to where the resonance pole resides, and to investigate how
this shape carries through to the three-body system. In Fig.~\ref{fig.6} we
give the cross section for the potentials TGE-B and SRW as examples of a
potential supporting a ``bound state'' and a ``virtual state'', respectively.
We observe that for TGE-B we have a classic resonance shape from which we
might be able to estimate the width of the resonance to be $\approx 5$~MeV.
However, for the SRW potential we have a sharp spike which could be
interpreted as a threshold effect.

To investigate how the shape of the cross section changes as the resonance pole
moves below the $\Sigma N$ threshold and approaches the real energy axis,
we have considered the potential TGE-B and varied the strength of the coupling
between the $\Lambda N$ and $\Sigma N$ channels.  We know the position of this
resonance pole, when the coupling is included, to be on the $[bt]$ sheet
at an energy of $(2131.7 - 5.4 i)$, which is just above the $\Sigma N$
threshold. This pole moves to $(2126.7-0 i)$ when the coupling is turned off.
This corresponds to a $\Sigma N$ bound state with a binding energy of
$4.3$~MeV. We introduce a new parameter $R$ in the coupling
\[
R \times C_{\Lambda\Sigma}
\]
and consider values of $R = 1.25, 1.0, 0.75, and 0.5$,  so that we can move the
position of the resonance pole from a point above the $\Sigma N$ threshold
to a point below the threshold and closer to the real axis. In
Fig.~\ref{fig.7}, we present the cross section for $\Lambda N$
scattering for the above values of $R$. We find, as expected, that as we
move the pole closer to the real axis $(R\rightarrow 0)$ the width of
the resonance is reduced. More important is the fact that, as we move below
the threshold and reduce the width, the shape of the resonance in the cross
section becomes more symmetric. This suggests that the $\Sigma N$ branch cut
has a shadowing effect on the cross section. A similar effect will be
observed for the three-body system.

\subsection{The Three-Body System}

We now turn to the $YNN$ system with the aim of examining the possible
formation
of $^{3}_{\Sigma}$H states near the $\Sigma NN$ threshold. If such states exist
for the two-body potentials under consideration, we expect to find them as
poles of the scattering amplitude in the complex energy plane, or as solutions
of the Schr\"odinger equation for complex energies. However, before we proceed
to explore the complex energy plane we should examine the quantum numbers such
a state would have. Considering the results of Dover and Gal~\cite{Dov82}, we
expect such states to have the lowest isospin possible for the $YNN$ system,
$T=0$. This suggests that we could observe these states in $\Lambda d$
scattering near the $\Sigma$ production threshold. Furthermore, because the
resonance in the $YN$ system occurs in the $S$-wave, we might expect the $YNN$
resonance to be in the $J^\pi=\frac{1}{2}^+$ channel. Thus, as a first step in
determining the possible existence of $\Sigma$-hypernuclear states, we
examine the total $S$-wave cross section in the $J^\pi=\frac{1}{2}^+$ partial
wave. We should remind the reader at this stage that a resonance will
appear in just one partial wave, which will determine the quantum numbers
of the resonant state.

{}From unitarity we can write the total cross section for $\Lambda d$
scattering
as
\begin{equation}
\sigma_T = \sum_{J^\pi}\ \sigma^{J^\pi}_T \ ,                   \label{eq:3.3}
\end{equation}
where the total cross section for the partial wave with total angular momentum
and parity $J^\pi$ is given in terms of the imaginary part of the partial
wave $T$-matrix by
\begin{equation}
\sigma^{J^\pi}_T = - \frac{4\pi^2\mu_{\Lambda d}(2J+1)}{k_0 (2s_\Lambda + 1)
(2s_d + 1)}\ \sum_{\cal L S}\ Im\left[T^{J^\pi}_{{\cal L S};{\cal L S}}\right]\
.                                                         \label{eq:3.4}
\end{equation}
Here, ${\cal S}$ and ${\cal L}$ are the the channel spin and orbital angular
momentum of the $\Lambda$ respectively, while $s_\Lambda$ and $s_d$ are
the spin of the $\Lambda$ and deuteron. The on-shell momentum is taken to be
$k_0$, and $\mu_{\Lambda d}$ is the reduced mass for the $\Lambda d$ system.
For $S$-wave we take ${\cal L}=0$ and therefore ${\cal S}=J$. In this case the
total elastic cross section can be written in terms of the $S$-wave amplitude
as
\begin{equation}
\sigma^{el}_T = \frac{4\pi^3\mu_{\Lambda d}^2}{(2s_\Lambda +1)(2s_d + 1)}
(2J+1)\ |T^{J^\pi}_{0J;0J}|^2     \ ,                      \label{eq:3.5}
\end{equation}
while the inelastic total cross section is given by the difference between the
total cross section and the total elastic cross section; {\it i.e.},
\begin{equation}
\sigma_T^{in} = \sigma_T - \sigma_T^{el} \ .          \label{eq:3.6}
\end{equation}
In Figs.~\ref{fig.8} - \ref{fig.11} we give the total cross section for the
$^3$S$_1$ potentials SRW, TGE-A, TGE-B, and TGE-C. The solid curve corresponds
to the total elastic cross section, while the dotted curve corresponds to the
total inelastic cross section. In general, any structure is more
pronounced in the inelastic cross section. Comparing the results for
the different potentials, we may conclude that the potential TGE-C has
marginal structure if any, while the others have more pronounced structure
just below the $\Sigma$ production threshold. The other general conclusion we
may draw is that the inelastic total cross section for potentials SRW and
TGE-B has a more symmetric shape than that for potential TGE-A. The important
question now is: does any of this structure in the cross section correspond to
a resonant state? That is, is it an eigenstate of the Hamiltonian for the $YNN$
system.

In Table~\ref{table4} we present the position of the poles of the amplitude
for the $YN$ and $YNN$ systems near the threshold for $\Sigma$ production.
Here we observe that potential TGE-C, which produced a very wide resonance
shape in the total inelastic cross section, does have a pole in the
amplitude on the $[bt]$ sheet; the half width of this resonance is
11~MeV which is consistent with what we would deduce from the cross section.
{}From an experimental point of view such a wide resonance would be hard to
observe, and to that extent will give little information on the structure of
the
Hamiltonian that generated the eigenstate. We also note that for potential
TGE-C
the two-body $YN$ resonance also lies far from the physical region. Although
the half width of the two-body resonance is only 5.3~MeV, the pole lies well
above the threshold for $\Sigma$ production.  The distance from the pole to
the physical region is more than the half width, due to the presence of the
branch cut which separates the resonance position and the physical region.
This branch cut actually shields the resonance from view in the physical
region.

We now turn to the potentials SRW (Fig.~\ref{fig.8}) and TGE-B
(Fig.~\ref{fig.10}). In this case the two-body system supports either a
``bound'' state or a ``zero energy'' bound state {\it when} the coupling
between the $\Lambda N$ and $\Sigma N$ is set to zero. Here we get a true
resonance for potential TGE-B with a half width of 8.9~MeV, which is
similar to the result for potential TGE-C, but because of the smaller half
width we observe more pronounced structure in the total cross section.
This width is comparable to that observed experimentally in the $A=4$
system. On the other hand, potential SRW gives a resonance with a half
width of 1.2~MeV and very pronounced structure.  In this case
the resonance is slightly above the threshold for $\Sigma$ production,
but because of its proximity to the physical region it has considerable
influence on the cross section. In this case, unlike the cross
section for potential TGE-B, the elastic total cross section falls sharply
at threshold, which suggests that for resonances on the $[bt]$ sheet and
above the $\Sigma$ production threshold the branch cut produces some shadowing
effect, similar to that seen in the $YN$ system for potential TGE-C.

Finally, for potential TGE-A we had some difficulty in determining the
actual position of the resonance pole. This, we think, was due to the fact that
the pole is very close to the $\Sigma$ production threshold, and as a result
our numerical procedures failed. (We performed the search with 64 point
Gauss Legendre points to convert the coupled integral equations to a
set of algebraic equations with no satisfactory convergence.) This numerical
problem is primarily due to the fact that no contour rotation can move the
$\Sigma NN$ branch point away from the integration path. The resonance position
that we list in Table~\ref{table4} is presented just to show that (i) the
resonance is very close to the threshold and (ii) as a result it produces a
rapid variation in the cross section over a small energy region near the
threshold. Here again, the structure in the cross section is not symmetric -- a
reflection of the fact that the resonance pole lies above the threshold for
$\Sigma$ production and the branch cut due to the threshold produces a
shadowing effect.

{}From a comparison of the results for the four potentials we may draw the
following conclusions regarding the correlation between the two- and
three-body systems (see Table~\ref{table4}). As the pole in the two-body
system moves from the $[tb]$ sheet to the $[bt]$ sheet, the width of the
resonance in the $YNN$ system increases. In other words, the presence of the
third baryon enhances the overall attraction in the system, effectively
``binding'' the $\Sigma NN$ system. When the situation is such that the
strength in the two-body interaction produces a pole close to the $\Sigma$
production threshold, then the pole in the three-body problem lies a
little farther from the corresponding $\Sigma$ production threshold. To
illustrate this point, we examine what happens when the interaction is
generated from potential TGE-B by modifying the
coupling between the $\Lambda N$ and $\Sigma N$ as described in the previous
section. In Fig.~\ref{fig.12} we present the total cross section for the
$J^\pi=\frac{1}{2}^+$ partial wave, as defined in Eq.~(\ref{eq:3.4}), for
$R=0.5,\,0.75,\,1.0$, and $1.25$.  By comparing the results in
Figs.~\ref{fig.7} and \ref{fig.12} we illustrate that,  as the pole in the $YN$
system moves closer to the real axis, the pole in the $YNN$ also moves closer
to
the physical region. However, the width of the resonance in the $YNN$ system,
as reflected in the total cross section, is in all cases larger than
that in the $YN$ system. The close relation between the result for the
$YN$ and $YNN$ systems indicates that we need to investigate, experimentally,
the cross section for $\Lambda p$ scattering near the $\Sigma$ production
threshold. This need, for more experimental information about the $\Lambda p$
cross section, is further bolstered by the fact that some of the OBE potential
models (which are fitted to the existing $\Lambda p$ data) exhibit resonance
type structure near or below the $\Sigma$ production threshold.

To demonstrate that the observed structure in the cross section is not a
threshold effect, we present in Fig.~\ref{fig.13} the partial--wave total cross
section $\sigma_T^{J^\pi}$ for the first four partial waves. Two important
conclusions can be drawn from the curves in this figure. First, the resonance
structure below the $\Sigma$ production threshold is observed only in the
$J^\pi=\frac{1}{2}^+$ partial wave. The other partial cross sections exhibit a
broad bump above the $\Sigma$ production threshold, which is due to the opening
of a new channel. In fact, this enhancement in the $J^\pi\ne \frac{1}{2}^+$
cross sections is a threshold effect that can be seen in all the non-resonant
partial waves, while the structure below the $\Sigma$ production lies only
in one partial wave allowing us to assign a definite quantum number to that
structure. The second interesting feature is that the $S$-wave cross section
is not the dominant contribution. The $P$-wave ($J^\pi=\frac{3}{2}^-$) total
cross section is larger. This is not unexpected considering the size of the
deuteron and the momentum of the incident $\Lambda$ at these energies.
Unfortunately, this will make it difficult to observe such
$\Sigma$-hypernuclear
resonances in $\Lambda d$ scattering, because the total cross section will be
dominated by non-resonant partial waves. We should recall that for $\Lambda p$
scattering it is the $S$-wave scattering that provides the main contribution to
the overall total cross section.  Thus, to observe a $\Sigma$-hypernuclear
state in the $A=3$ system, one must consider reactions that can select, or
enhance, the ($T=0$, $J^\pi=\frac{1}{2}^+$) channel.

\section{CONCLUSIONS}

Using separable $NN$ and $YN$ potentials in the Faddeev equations for the
YNN system, we have demonstrated that the structure in the model $\Lambda d$
cross section near the $\Sigma NN$ threshold is associated with resonance
poles in the scattering amplitude.  The positions of these poles on the
second Riemann sheet of the complex energy plane are, in fact, the eigenvalues
of the analytic continuation of the kernel of the Faddeev equations.  Perhaps
surprisingly, the cut starting at the $\Sigma NN$ threshold appears to
shield from view in the physical region those resonance (pole) singularities
lying above that threshold.  Therefore, whether the resonance pole,
corresponding to a $\Sigma NN$ eigenstate, lies above or below the
$\Sigma NN$ threshold, the structure appearing in the $\Lambda d$ cross
section lies {\it below} the $\Sigma NN$ threshold.  If the pole resides
below the $\Sigma NN$ threshold, then the structure in the cross section
takes the shape of a classic resonance, symmetric about the real part of
the resonance eigenvalue.  In contrast, for a pole that lies in
the shadow of the $\Sigma NN$ cut, the structure can be quite distorted,
falling sharply at threshold and producing a more cusp-like shape.  In such
a case, the position of the peak in the structure does not necessarily
correspond to the real part of the resonance eigenvalue, because the pole
position is shielded from view in the physical region.  Clearly, any shape
intermediate between these two extremes is possible, so that one cannot
necessarily determine whether a pole lies above or below the $\Sigma NN$
threshold from the shape of the resonance structure in the $\Lambda d$ cross
section.  Nonetheless, structure below the $\Sigma NN$ threshold in the
$\Lambda d$ cross section, like that which has been observed in the
$^4$He(K$^-,\pi^-$) reaction, does imply the existence of a resonance
(an eigenstate of the Hamiltonian in a particular partial wave) in the
$\Sigma NN$ system.

That the cross section structure in the model $\Lambda d$ scattering
calculation is a resonance and not just a threshold effect was established
by demonstrating that the structure lies only in the $\frac{1}{2}^+$ partial
wave, and not in the neighboring channels.  Unfortunately, the $L=0$
partial wave does not dominate the $\Lambda d$ cross section, as is the
case in $\Lambda p$ scattering.  Therefore, to observe a $\Sigma$-hypernuclear
state in the A=3 system, one must consider reactions that can select, or
enhance, the $\frac{1}{2}^+$ channel.

Finally, in the hypertriton the presence of three baryons enhances the
attraction in the unbound $\Lambda N$ system, such that the $\Lambda NN$
system is bound with respect to separation of the $\Lambda$ from the
deuteron.  Similarly, the presence of the second nucleon enhances the
overall attraction in the $\Sigma NN$ system, effectively ``binding''
that system to produce a resonance pole.  Furthermore, we found that, as
the pole in the $YN$ system moves closer to the real axis, the pole in
the $YNN$ system moves closer to the physical region.  However, the width
of the resonance in the $YNN$ system is always larger than that in the
$YN$ subsystem.

\acknowledgements

The work of I.~R.~Afnan was supported by the Australian Research Council.
That of B.~F.~Gibson was performed under the auspices of the U.~S.~Department
of Energy.  The authors thank B.~C.~Pearce for assistance in determining the
positions of the poles of the $YN$ amplitudes and S.~B.~Carr for help
in calculating the $\Lambda N$ cross sections.

\newpage

\unletteredappendix{Formal Solution of the AGS Equations}

In this appendix we present a formal solution of the integral equation for the
three-particle scattering amplitude in terms of the eigenstates  of the kernel
of the corresponding homogeneous integral equation. In this way we establish
the
relation between the poles of the scattering amplitude on the second Riemann
sheet of the energy plane, and the eigenstates of the Hamiltonian for the
three-body system.

Let us consider the AGS equation for the amplitude $X_{\alpha\beta}$ as given
in
Eq.~(\ref{eq:2.10}),
\begin{equation}
X_{\alpha\beta} = G_0(E)\,\bar{\delta}_{\alpha\beta} +
\sum_\gamma\,G_0(E)\,\bar{\delta}_{\alpha\gamma}\,T_\gamma(E)\,
X_{\gamma\beta} \ .                                   \label{eq:A.1}
\end{equation}
The corresponding homogeneous equation is given by
\begin{equation}
|\phi_\alpha\,\rangle =
\sum_\beta\,G_0(E)\,\bar{\delta}_{\alpha\beta}\,T_\beta\,
|\phi_\beta\,\rangle       \ .                        \label{eq:A.2}
\end{equation}
This equation is basically the Schr\"odinger equation for the three-body
system, and the determination of the energies at which this equation is
satisfied gives us the spectrum of our three-body Hamiltonian. Thus, any
solutions of this equation for negative real energies correspond to bound
states.  To determine the position of the resonance poles which are not on the
first Riemann sheet of the complex energy plane, we need to extend the energy
domain of Eq.~(\ref{eq:A.2}). This can be achieved in momentum space by
deforming the contour of integration such that
$q\rightarrow q\,e^{-i\theta}$~\cite{Afn91}, where $\theta$ is the angle of
rotation of the integration variables, in this case
the momentum. In this way we have extended the energy domain over which
Eq.~(\ref{eq:A.2}) is defined to that part of the second Riemann sheet where
resonances are normally located. The resulting equation is denoted by
\begin{equation}
|\phi^\theta_\alpha\,\rangle = \sum_\beta\,G_0^\theta(E)\,
\bar{\delta}_{\alpha\beta}\,T^\theta_\beta(E)\,|\phi^\theta_\beta\,\rangle \ .
                                                         \label{eq:A.3}
\end{equation}
Here, the energy $E$ can be in that part of the second Riemann sheet where the
$\arg(E) > - 2\theta$. In general, there are limitations on this deformation
of the contour imposed by the singularities of the kernel. This limitation
puts a constrain on the resonances that can be studied using this approach.
To solve Eq.~(\ref{eq:A.3}), we need to consider the corresponding eigenvalue
problem,
\begin{equation}
\lambda_n(E)\,|\phi^\theta_{n,\alpha}\,\rangle = \sum_\beta\ G_0^\theta(E)\,
\bar{\delta}_{\alpha\beta}\,T_\beta^\theta(E)\,|\phi^\theta_{n,\beta}\,\rangle
\ ,                                                      \label{eq:A.4}
\end{equation}
where $\lambda_n$ is the eigenvalue of the kernel of the three-body integral
equation. For those energies for which there is an eigenvalues, $\lambda_n(E)$,
whose value  is one, Eq.~(\ref{eq:A.3}) is said to have a solution. This
solution is an eigenstate of the full three-body Hamiltonian, even when the
energy $E$, is complex provided it is on the second Riemann sheet.


To expand the three-particle scattering amplitude $X_{\alpha\beta}(E)$ in
terms of the solutions of Eq.~(\ref{eq:A.4}) ({\it i.e.}, the eigenvectors
of the kernel of the integral equation) we must determine the orthonormality
condition for the eigenstates $|\phi^\theta_{n,\alpha}\,\rangle$. For this
we need to introduce the eigenvalue equation for the case when the rotation
of the contour of integration is taken to be $q\rightarrow q\,e^{i\theta}$,
and the resultant equation is
\begin{equation}
\tilde{\lambda}_n(E)\,|\tilde{\phi}^\theta_{n,\alpha}\,\rangle =
\sum_\beta\ G_0^{-\theta}(E)\,\bar{\delta}_{\alpha\beta}\,
T_\beta^{-\theta}(E)\,|\tilde{\phi}^\theta_{n,\beta}\,\rangle \ .
                                                        \label{eq:A.5}
\end{equation}
This equation extends the energy domain of our eigenvalue problem to that part
of the second Riemann sheet where the solutions of the adjoint kernel reside.
Making use of the fact that the kernels of Eqs.~(\ref{eq:A.4}) and
(\ref{eq:A.5}) are related by
\begin{equation}
\left[T^{-\theta}_\alpha(E^*)\,G^{-\theta}_0(E^*)\right]^* =
T^\theta_\alpha(E)\,G^\theta_0(E)     \ .                \label{eq:A.6}
\end{equation}
we can show that the eigenstates of the homogeneous equation satisfy the
orthonormality condition
\begin{equation}
\sum_\alpha\
\langle\tilde{\phi}^\theta_{m,\alpha}(E^*)\,|\,T^\theta_\alpha(E)\,
|\,\phi^\theta_{n,\alpha}(E)\,\rangle = \delta_{nm} \ .   \label{eq:A.7}
\end{equation}
Here we note that unlike bound state solutions, the normalization involves the
eigenstates of the kernel and the adjoint kernel.  Because the kernel is not
Hermitian, which was the case for bound state, we state the orthonormality
of the resonance wave function in terms of two eigenvalue equations.

We are now in a position to expand the scattering amplitude
$X_{\alpha\beta}(E)$ in terms of the eigenstates
$|\phi_{n,\alpha}(E)\,\rangle$.
In particular, if we want the amplitude on that part of the second Riemann
sheet where the resonance poles reside, we must write the expansion in terms of
the eigenstates of the rotated kernel; {\it i.e.},
\begin{equation}
X^\theta_{\alpha\beta}(E) = \sum_n\ |\,\phi^\theta_{n,\alpha}\,\rangle\,
C_{n\beta}(E)\ .                                           \label{eq:A.8}
\end{equation}
The constants $C_{n\beta}(E)$ can be determined by substituting the expansion
in Eq.~(\ref{eq:A.8}) in the integral equation for the scattering amplitude on
the rotated contour, Eq.~(\ref{eq:A.1}) on the rotated contour. This
gives us an expansion for the scattering amplitude in terms of the eigenstates
and
eigenvalues of the kernel of the integral equation of the form
\begin{equation}
X_{\alpha\beta}(E) = \sum_n\ |\,\phi^\theta_{n,\alpha}(E)\,\rangle\,
\frac{\left[\tilde{\lambda}_n(E^*)\right]^*}{1 - \lambda_n(E)}\,
\langle\,\tilde{\phi}^\theta_{n,\beta}(E^*)\,| \ , \label{eq:A.9}
\end{equation}
In writing Eq.~(\ref{eq:A.9}), we have established the fact that the energy at
which one of the eigenvalues $\lambda_n(E)$ is one, the amplitude
$X_{\alpha\beta}(E)$ has a pole. However, the energies at which the eigenvalues
are one, correspond to solutions of the homogeneous Eq.~(\ref{eq:A.3}), which
correspond to eigenstates of the Hamiltonian when the energy domain on which
this Hamiltonian is defined is extended onto the second Riemann sheet. In
this way we have established the fact that; poles of the scattering amplitude
on the second Riemann sheet of the energy plane, which correspond to
resonances, are also the positions of the eigenstates of the Hamiltonian
when the energy domain is extended to the second Riemann sheet.

\newpage

\figure{The branch cuts and thresholds in the complex energy
plane.\label{Fig.1}}

\figure{Region of the energy plane that can be accessed via contour rotation.
The point $A$ at $E=M_r-\frac{i}{2}\Gamma + m_\alpha$ corresponds to the
branch point resulting from the resonance in the two-body
subsystem.\label{Fig.2}}

\figure{The shaded area is the domain of the second Riemann sheet of the
energy plane to which we can analytically continue Eq.~(\ref{eq:2.17}),
while maintaining the contour deformation along a ray in the fourth quadrant
of the $q'$-plane.\label{Fig.3}}

\figure{The shaded area is the domain of the third Riemann sheet of the energy
plane to which we can analytically continue Eq.~(\ref{eq:2.17}), while
maintaining the contour deformation along a ray in the fourth quadrant of
the $q'$-plane. Access to this sheet is via the square root branch cut
resulting from the resonance pole in $\tau_{\kappa_\alpha}$.\label{Fig.4}}

\figure{The labeling of the different Riemann sheets for the $\Lambda
N$-$\Sigma
N$ coupled-channel problem. This labeling scheme is identical to that used in
Ref.~\cite{Pea89}.\label{Fig.5}}

\figure{The total cross section for $\Lambda N$ scattering in the $^3$S$_1$
channel for the two potentials SRW and TGE-B.\label{fig.6}}

\figure{The total cross section for $\Lambda N$ scattering in the $^3$S$_1$
channel for the TGE-B potential with the coupling between the $\Lambda n$ and
$\Sigma N$ channels $C_{\Lambda\Sigma}$ replaced by $R \times
C_{\Lambda\Sigma}$.\label{fig.7}}

\figure{The total elastic (solid line) and inelastic (dotted line) $S$-wave
$J^{\pi}=\frac{1}{2}^+$ cross section for $\Lambda d$ scattering as a function
of the three-body energy for the potential SRW. The $\Sigma NN$ Threshold is at
$E_{cm}=77$~MeV.\label{fig.8}}

\figure{The total elastic (solid line) and inelastic (dotted line) $S$-wave
$J^{\pi}=\frac{1}{2}^+$ cross section for $\Lambda d$ scattering as a function
of the three-body energy for the potential TGE-A. The $\Sigma NN$ threshold is
at $E_{cm}=77$~MeV.\label{fig.9}}

\figure{The total elastic (solid line) and inelastic (dotted line) $S$-wave
$J^{\pi}=\frac{1}{2}^+$ cross section for $\Lambda d$ scattering as a function
of the three-body energy for the potential TGE-B. The $\Sigma NN$ threshold is
at $E_{cm}=77$~MeV.\label{fig.10}}

\figure{The total elastic (solid line) and inelastic (dotted line) $S$-wave
$J^{\pi}=\frac{1}{2}^+$ cross section for $\Lambda d$ scattering as a function
of the three-body energy for the potential TGE-C. The $\Sigma NN$ Threshold is
at $E_{cm}=77$~MeV.\label{fig.11}}

\figure{The total cross section in the $J^{\pi}=\frac{1}{2}^+$ partial wave for
potential TGE-B with the coupling strength between the $\Lambda N$ and $\Sigma
N$ scaled by the factor $R$; {\it i.e.}, $C_{\Lambda\Sigma}\rightarrow R\times
C_{\Lambda\Sigma}$.\label{fig.12}}

\figure{The total cross section for partial waves $J^\pi= \frac{1}{2}^+,
\frac{1}{2}^-, \frac{3}{2}^+, \frac{3}{2}^-$.\label{fig.13}}

\newpage

\begin{table}

\caption{The parameters of the $^3$S$_1$ $\Lambda N$-$\Sigma N$ coupled-channel
potentials, and the $^1$S$_0$ SRW potential.\label{table1}}

\vskip 0.3 cm

\begin{tabular}{l|ccccc}
Potential & $C_{\Lambda\Lambda}$ & $\beta_\Lambda$ & $C_{\Sigma\Sigma}$ &
$\beta_\Sigma$ & $C_{\Lambda\Sigma}$ \\ \hline
SRW $^3$S$_1$   &-0.42824 & 1.6    &-1.88913 & 2.0    &0.84289 \\
TGE-A           &-0.11729 & 1.1069 &-4.33140 & 2.702  &0.71399 \\
TGE-B           & 0.03569 & 0.9518 &-0.80233 & 1.2789 &0.43692 \\
TGE-C           & 0.05726 & 0.8752 &-0.07434 & 0.5335 &0.23226 \\
SRW $^1$S$_0$   &-0.17339 & 1.18   & 0.45856 & 1.44   &-0.38471 \\
\end{tabular}
\end{table}

\vskip 1cm

\begin{table}
\caption{The position of the poles of the $\Lambda N-\Sigma N$ amplitude that
lie close to the $\Sigma N$ threshold for the four different $^3$S$_1$ $YN$
interaction being considered.$^{\rm a}$\label{table2}}
\begin{tabular}{l|ccc}
Potential & Sheet & Pole with & Pole with \\
 & & $C_{\Lambda\Sigma}\ne 0$ & $C_{\Lambda\Sigma}=0$ \\ \hline
SRW    & [tb] & $2132.5 - 0.4 i$ & $2131.0 + 0.0 i$ \\
TGE-A  & [tb] & $2130.9 - 1.9 i$ & $2130.5 + 0.0 i$ \\
TGE-B  & [bt] & $2131.7 - 5.4 i$ & $2126.7 + 0.0 i$ \\
TGE-C  & [bt] & $2138.0 - 5.3 i$ & $2129.0 + 0.0 i$ \\
\end{tabular}
\tablenotes{$^{\rm a}$Here, and throughout this paper we have taken our masses
to be $m_N=939$~MeV, $m_\Lambda=1115$~MeV, and $m_\Sigma= 1192$~MeV.
As a result the threshold for $\Sigma$ production is $2131$~MeV.}
\end{table}
\vskip 1 cm

\begin{table}
\caption{The effective range parameters for the $\Lambda N-\Sigma N$
coupled-channels in the $^3$S$_1$ partial wave. We have included both the
$\Lambda N$ and $\Sigma N$ effective range parameters.\label{table3} }

\begin{tabular}{l|cccc}
Potential & $a_{\Lambda N}$ & $r_{\Lambda N}$ & $a_{\Sigma N}$ & $r_{\Sigma N}$
\\ \hline
SRW $^3$S$_1$   & $-1.96$ & $2.44$ & $ 0.14 - 4.72 i$ & $ 1.67 - 0.20 i$ \\
TGE-A           & $-2.46$ & $3.94$ & $-2.60 - 2.97 i$ & $ 1.30 - 0.04 i$ \\
TGE-B           & $-1.70$ & $4.55$ & $ 2.97 - 1.83 i$ & $ 1.97 - 0.38 i$ \\
TGE-C           & $-1.69$ & $4.88$ & $ 3.81 - 1.56 i$ & $ 2.80 - 1.88 i$ \\
SRW $^1$S$_0$   & $-1.98$ & $4.03$ & $ 0.59 - 0.09 i$ & $-1.30 - 0.39 i$ \\
\end{tabular}
\end{table}

\vskip 1 cm

\begin{table}
\caption{The position of the poles of the $YNN$ amplitude near the $\Sigma NN$
threshold. Included are also the position of the resonance pole in the $YN$
amplitude for comparison.$^{\rm a}$\label{table4}}

\begin{tabular}{l|cc|cc}
            &\multicolumn{2}{c|}{Two-Body}&\multicolumn{2}{c}{Three-Body}\\
Potential & Sheet & Position & Sheet & Position \\ \hline
SRW       & $[tb]$  & $78.5 - 0.4 i$ & $[bt]$  & $79.5 - 1.2 i$ \\
TGE-A     & $[tb]$  & $76.9 - 1.9 i$ & $[bt]$  & $78 - 0.5 i\ ^{\rm b}$ \\
TGE-B     & $[bt]$  & $77.7 - 5.4 i$ & $[bt]$  & $75.5 - 8.9 i$ \\
TGE-C     & $[bt]$  & $84.0 - 5.3 i$ & $[bt]$  & $84.0 - 11.0 i$ \\
\end{tabular}
\tablenotes{$^{\rm a}$The energy of the two-body resonance is taken relative to
the $\Lambda N$ threshold, $2054$~MeV.}
\tablenotes{$^{\rm b}$Because the pole position is close to the $\Sigma NN$
threshold, we found it difficult to determine the position of the pole with a
high degree of accuracy}
\end{table}

\end{document}